%% file: main.tex
\documentclass[submission,copyright,creativecommons,noderivs,noncommercial]
              {eptcs}

\usepackage[T1]{fontenc}
\usepackage{microtype}

\usepackage{amsmath}
\usepackage{amssymb}
\usepackage{amsthm}

\newtheorem{definition}{Definition}

\usepackage{paralist}

\usepackage{tabularx}
\usepackage{booktabs}

\usepackage{tikz-cd}
\usepackage{subfig}

\usepackage{float}

\usepackage{hyperref}
\hypersetup{
  colorlinks=true,
  allcolors=blue,
  linkcolor=black
}

\usepackage{booktabs}

\newcommand{\join}{\cup}
\newcommand{\partof}{\mathop{\subseteq}}

\newcommand{\liff}{\leftrightarrow}

\newcommand{\system}[1]{\ensuremath{\mathrm{#1}}}
\newcommand{\axiomdef}[1]{\raisebox{\ht\strutbox}{\hypertarget{#1}{}}{\ensuremath{\mathbf{#1}}}}
\newcommand{\axiom}[1]{\hyperlink{#1}{\ensuremath{\mathbf{#1}}}}

\makeatletter
\def\modal{\@ifnextchar<{\pos@internal}{\@ifnextchar[{\nec@internal}{}}}
\def\pos@internal<#1>{%
  \if\relax\detokenize{#1}\relax%
  \pos%
  \else%
  {\langle#1\rangle}
  \fi%
}
\def\nec@internal[#1]{%
  \if\relax\detokenize{#1}\relax%
  \nec%
  \else%
  {[#1]}%
  \fi
}
\def\modalup{\@ifnextchar<{\posup@internal}{\@ifnextchar[{\necup@internal}{}}}
\def\necup@internal[#1]{\modal[{\uparrow}#1]}
\def\posup@internal<#1>{\modal<{\uparrow}#1>}
\def\modaldown{\@ifnextchar<{\posdown@internal}{\@ifnextchar[{\necdown@internal}{}}}
\def\necdown@internal[#1]{\modal[{\downarrow}#1]}
\def\posdown@internal<#1>{\modal<{\downarrow}#1>}
\makeatother

\def\::{\mathpunct:}

\mathchardef\breakingcomma\mathcode`\,
{\catcode`,=\active
  \gdef,{\breakingcomma\discretionary{}{}{}}
}
\newcommand{\mathlist}[1]{\mathcode`\,=\string"8000 #1}

\providecommand{\event}{FMAS 2021}
\title{Towards a Formalisation of Justification and Justifiability%
  \thanks{This work is partly funded by the German Research Council (DFG) 
  as part of the project
  ``Integrated Socio-technical Models for Conflict Resolution
  and Causal Reasoning'', grant no. FR 2715/5-1, DA 206/12-1.%
}}

\author{Willem Hagemann
  \institute{Department of Computing Science\\University of Oldenburg\\Germany}
  \email{willem.hagemann@uol.de}
}
\def\titlerunning{Towards a Formalisation of Justification and Justifiability}
\def\authorrunning{Willem Hagemann}

\makeatletter
\hypersetup{%
    pdfencoding=auto,
    pdfauthor={\authorrunning},
    pdftitle={\titlerunning}
}
\makeatother

\allowdisplaybreaks
\DeclareMathAlphabet{\mathcal}{OMS}{cmsy}{m}{n}

\begin{document}
\maketitle
\input{abstract}
\input{introduction}
\input{just_logic}

\input{more_logic}

\input{conclusion}

\bibliographystyle{eptcs}
\bibliography{bibliography}

\end{document}

%% file: abstract.tex
\begin{abstract}
  We introduce the logic $QK^D_{\mathcal S}$ which is a normal multi-modal
  logic over finitely many modalities that additionally
  supports bounded quantification of modalities.
  An important feature of this logic is that it allows to quantify over
  the information components of systems and, hence, can be used to
  derive justifications. We compare the proposed logic with
  Artemov's justification logic
  and also report on a prototypical implementation of a satisfiability
  solver of this logic and show some examples.
\end{abstract}

%% file: introduction.tex
\section{Introduction}
We are in the process of building complex autonomous systems
that perceive their environment, exchange information and decide
on their actions based on their understanding of the world.
The hope is that the multitude of perspectives will lead to a more
complete view of the world and enable systems to act more optimally.
The problem is that information from different sources is often
inconsistent. Inconsistencies are one of the major concern of formal methods,
as the presence of an inconsistency usually means that a contradiction
is derivable, and thus\,---\,ex falso quodlibet\,---\,any proposition
is derivable.

A wide variety of formal methods have been developed for the analysis and verification of complex cyber-physical systems (CPS).
A central approach is to create a formal model of the system
(and its environment) and then to prove certain properties
using rigorous logical-algorithmic methods such as model checking \cite{Alur1995,Damm2012,clarke2018model}. However, model checking of cyber-physical systems
is often limited to the verification of temporal properties.
Common assumptions are that the internal state and control laws of the system
are precisely known and that such systems are deployed in limited
environmental scenarios. Hence, the temporal behaviour of the system and
its environment can be predicted within quantifiable uncertainties \cite{Frehse2011}.

Especially with regard of autonomous CPS, such simplifying assumptions must
be revised or\,---\,at least\,---\,be re-examined
\cite{platzer2019logical,baier2021verification}.
Autonomous systems are supposed to operate in complex and open environments with a high degree of independence
and self-determination. An autonomous system knows its action alternatives
and independently decides for one of the possible actions \cite{fisher2013verifying}.
Such a decision, however, shall then be justifiable and explainable\,---\,a requirement that 
research projects like XAI \cite{XAI} or Perspicuous Computing \cite{CPEC}
address. Moreover, a justification shall run through the entire perception and
decision-making process of the system. Ideally, a justification keeps record
of the sources from which the information originates
that was used for decision-making.
Such sources of information are, for example, external agents,
but also internal subsystems, such as basic control laws, neural networks for
image recognition/classification \cite{platzer2019logical},
or a catalogue of traffic rules \cite{jsan10030041}.

Increasingly, modal logic concepts are proposed to formally describe autonomous
systems as epistemic agents
\cite{platzer2019logical,dissing2020implementing,fisher2013verifying}.
The numerous epistemic logics dealing with belief and knowledge
(as true belief) in multi-agent systems \cite{fagin2003,van2007dynamic}
are joined by other modal logics.
E.g., the Center for Perspicuous Computing \cite{CPEC} proposes
to use the knowledge representation of description logics \cite{10.1007/11814771_26,sebastiani2009automated,HHMW12}, whereas \cite{platzer2019logical} proposes to use differential dynamic logics, and \cite{fisher2013verifying} refers to the belief--desire--intention framework \cite{rao1995bdi}.
Another important candidate is Artemov's
justification logic \cite{artemov:logic-justification:2008}, as this
logic promises for the first time to add a justification component to
previous epistemic logics, thus satisfying Plato's characterisation
of knowledge as justified true belief.

This paper deals with justification in a fallibilist process of cognition and decision making. Instead of belief and knowledge, it is based on the weaker
notion of information, see Barwise \cite[Ch. 9]{barwise1989situation} and Gerbrandy \cite{gerbrandy1998} for a short discussion of the contrast of knowledge
and information. Unlike belief or knowledge, information need not be
consistent in its entirety. Consequently, we abandon destructive techniques
that contract the current set of beliefs as proposed in the AGM framework\footnote{The framework is named after its proponents, Alchourrón, Gärdenfors, and Makinson.} of belief revision
\cite{van2007dynamic}. Rather, we stockpile and maintain each source of
information regardless of whether or not it contributes to a consistent
worldview. A statement is justifiable in this setting if and only if
we find a group of information sources that is free of
contradiction in its totality and from which the statement can be substantiated.

In Sec.~\ref{sec:motivation}, I propose to use the
well-known multimodal logic $K^D_{\mathcal S}$\,---\,also referred to as $K^D_n$
in \cite{fagin2003}\,---\,%
to model the information distribution within an information processing system
$\mathcal S$, where $\mathcal S=\{s_1,\dots,s_n\}$ is a the set of atomic
information components $s_1$, \dots, $s_n$ of a system.
In $K^D_{\mathcal S}$ the \emph{principle of information distribution}%
\,---\,referred to as \emph{distributed knowledge} in \cite{fagin2003}\,---\,%
holds. That is, we consider arbitrary nonempty subsets of $\mathcal S$,
called instances of $\mathcal S$. The information of an instance
is given as the deductive closure of the information of its atomic components.
Hence, $K^D_{\mathcal S}$ allows us to reason about the information of
arbitrary combinations of information components. This in turn yields
an analogous notion of justifications as in Artemov's justification logic
\cite{artemov:logic-justification:2008,artemov:jck:2006}. 
%
I gradually introduce basic features
of $K^D_{\mathcal S}$ along an application example of automated driving.
Besides proposing $K^D_{\mathcal S}$ as a justification logic
and comparing it with Artemov's justification logic, this section does
not contain any new results, but compiles those results which will be used
in the rest of the paper.

In Sec.~\ref{sec:justifiability}, I propose the extension of $K^D_{\mathcal S}$
by quantification over instances yielding the logic $QK^D_{\mathcal S}$, and
propose a tableau calculus for $QK^D_{\mathcal S}$. With the help of quantifiers,
the justifiability of a statement $\phi$ is defined formally as
an existential quantified formula running over consistent instances
of the systems $\mathcal S$.
Afterwards, I revisit the running example and show some results
with the help of a prototypical implementation of this tableau.

Finally, in Sec.~\ref{sec:conclusion}, I summarise the presented work,
and address a still outstanding comparison of the solver with other solvers.
I also address possible extensions of $QK^D_{\mathcal S}$ and mention conceivable philosophical positions that are
touched by this approach.

%% file: just_logic.tex
\section{A Normal Logic of Justification}\label{sec:motivation}
The following running example will guide us to through this section.
A vehicle with an automated driving system supporting conditional
driving automation (SAE Level 3, \cite{sae}) is driving on a multi-lane road.
The vehicle observes that $\modal[o_1]$ the current lane is blocked 
and that $\modal[o_2]$ there is a gap on the neighbouring lane that
is just large enough for a lane change.
The car has to decide whether it should brake or use the small gap for the
lane change. In addition to the observation data, the driving system of the
vehicle uses the following information in its decision-making process:
\begin{itemize}
\item The driving system adheres to the following rules:
  $\modal[r_1]$ Whenever the lane is blocked, brake or change lane.
  $\modal[r_2]$ Never change lane when no gap is available.
  $\modal[r_3]$ Never brake when the goal is to drive fast.
  $\modal[r_4]$ When the goal is to drive considerately, then change lane
  only if there is a large gap.
\item Since the user is in a hurry, for this trip he set the goal $\modal[g_1]$
  to drive fast. During the initial start-up dialog, the user has chosen
  $\modal[g_2]$ a considerate driving behaviour as default.
\end{itemize}
Let $\Sigma$ be a formalisation\footnote{A hint on the formalisation of
rule $\modal[r_4]$ seems appropriate. The rule consists of two nested conditionals, the
latter being ``change lane only if there is a large gap''.
The correct formalisation of the ``only if''-expression
is $\texttt{changeLane} \to \texttt{largeGapAvailable}$, just as
the popular mathematical jargon ``$A$ if and only if $B$'' decomposes
into ``$A$ if $B$'', formalised as $B \to A$, and ``$A$ only if $B$'', formalised as $A\to B$. Hence, the whole rule is formalised as $\texttt{beConsiderate} \to (\texttt{changeLane} \to \texttt{largeGapAvailable})$. Finally, the listed form is obtained by a simple equivalence transformation.} of all given information as follows:
\begin{align*}
  &\modal[o_1] \texttt{laneBlocked}, \qquad \modal[o_2] (\texttt{gapAvailable} \land \lnot\texttt{largeGapAvailable}),\\
  &\modal[r_1] (\texttt{laneBlocked} \to \texttt{brake} \lor \texttt{changeLane})\\
  &\modal[r_2]( \texttt{changeLane} \to \texttt{gapAvailable})\\
  &\modal[r_3]( \texttt{beFast} \to \lnot\texttt{brake})\\
  &\modal[r_4]( \texttt{beConsiderate} \land \texttt{changeLane} \to
  \texttt{largeGapAvailable})\\
  &\modal[g_1] \texttt{beFast}, \qquad \modal[g_2]\texttt{beConsiderate}.
\end{align*}
The given information is contradictory as there
is no satisfying truth assignment of the propositional variables of the
formalisation. Hence, the automated driving system
sends out a request to intervene to the user.
The surprised user takes control, gets an overview but misses to use
the small gap. Since this situation occurs several times during his trip,
he finally misses his appointment and sends a disappointed experience
report to the manufacturer. 

\subsection{Labels are Normal Modalities}
We put ourselves in the role of the manufacturer, analyse the
formalisation of the situation described above, and immediately identify
the unsatisfiability of $\Sigma$ as a cause of the request to intervene.
Since we already structured the information by labels, it is quite natural
to ask what labels contribute to the contradiction.

To this end, let us discuss the role
of the labels first. The labels refer to various information components.
For example, $\modal[g_1]$ refers to the initial start-up dialog of the car,
$\modal[r_4]$ refers to the engineer who added comfort functions
to the vehicle, and $\modal[o_1]$ refers to the front radar of the vehicle.
The notation $\modal[o_1]\phi$ denotes that $\phi$ belongs to the information
of $\modal[o_1]$, and\,---\,contrary to the given situation above\,---\,%
$\modal[o_1]\lnot\texttt{laneBlocked}$ describes a situation
where the front radar has the information that the lane is not
blocked. Moreover, we use negation of labelled formulae
to denote the absence of information:
$\lnot\modal[o_1]\texttt{laneBlocked}$ formalises that the front radar
does not have the information that the lane is blocked, i.e., the front radar
considers it possible that the lane is not blocked. Often, we
write $\modal<o_1>\phi$ instead of $\lnot\modal[o_1]\lnot\phi$.
Readers familiar with modal logic immediately recognise that $\modal[o_1]$ and
$\modal<o_1>$ are a pair of dual modal operators. In the following,
any label $s$ will be a modality and $\modal[s]$ the corresponding
normal modal operator. A normal modal operator $\modal[s]$ has access
to all logical tautologies via the necessitation rule
\begin{gather*}
  \text{from $\vdash \phi$ infer $\vdash \modal[s]\phi$,}\tag{\axiom{Nec}}
\end{gather*}
and its information is closed under logical consequence. The latter is obtained by the Kripke schema
\begin{gather*}
  \vdash \modal[s]( \phi \to \psi) \to \modal[s] \phi \to \modal[s]\psi.\tag{\axiom{K}}
\end{gather*}
We denote the set of all labels by $\mathcal S$ and we will refer to elements of
$\mathcal S$ as atomic modalities, or\,---\,for short\,---\,as atoms.

\subsection{The Principle of Information Distribution}
In order to identify those atoms of $\mathcal S$
that contribute to a contradiction, it is advantageous to consider
arbitrary \emph{nonempty} sets of atomic modalities, to which in turn we also
assign modal operators. We will refer to such sets as instances. E.g.,
the instance $\{o_1,o_2\}$ consists of the atoms $o_1$ and $o_2$ and
$\modal[o_1,o_2]$ denotes its assigned normal modal operator.
Further, we stipulate that for all instances the following principle
of \emph{information distribution}\,---\,in \cite{fagin2003} referred to as \emph{distributed knowledge}\,---\,holds:
For all instances $s=\{s_1,\dots,s_n\}\subseteq\mathcal S$ and
$t=\{t_1,\dots,t_m\}\subseteq\mathcal S$ it holds
\begin{gather*}
  \vdash \modal[s]( \phi \to \psi) \to \modal[t] \phi \to \modal[ s,t ]\psi\tag{\axiomdef{Dist'}}
\end{gather*}
where $s,t=\{s_1,\dots,s_n,t_1,\dots,t_m\}$.\footnote{Note
that we deliberately avoid the obvious use of the set operation $s \join t$ and prefer to write $s,t$ here, as this would yield confusion with
the reversed notational conventions in Boolean modal logic or description logic
$\mathcal{ALB}$. Indeed, in these logics the proper notation of $s,t$ would
be rendered by $s\cap t$.}
Hence, $[o_1,o_2]$ has access to all information that $o_1$ or $o_2$ have,
and also to all logical consequences that can be obtained by combining the
information of $o_1$ and $o_2$, as $[o_1,o_2]$ is again closed under logical
consequence. 

\subsection{Syntax and Axiomatisation of \texorpdfstring{$K^D_{\mathcal S}$}{KDS}}
We formally introduce the logic $K^D_{\mathcal S}$. It turns out that
$K^D_{\mathcal S}$ is actually a normal multi-agent logic with an axiom for
information distribution over (nonempty) groups of agents.
The axiom of information distribution (or an equivalent form thereof)
is often found in multi-agent epistemic logics, sometime referred to as
group information or distributed knowledge
\cite{halpern1985,fagin2003,gerbrandy1998}.
However, the presented logic lags behind most epistemic logics,
as it does not make any stronger requirements on the modality
of information. Information neither has to be factually true, like it is
the case for epistemic logics dealing with knowledge, nor be free
of contradictions, as it is usually stipulated in logics of
belief. The main difference lies in the designation of the modalities. While
the usual epistemic logics identify the modalities with agents, we
use modalities for fairly liberally chosen information sources.

The following two definitions give the language
and an axiomatisation of $K^D_{\mathcal S}$.
\begin{definition}
  Let $\mathcal V$ be a enumerable set of propositional variables and
  $\mathcal S$ be a finite set of atomic modalities.
  A formula $\phi$ of $K^D_{\mathcal S}$ is any expression that satisfies the
  Backus-Naur form
  \begin{gather*}
    \phi ::= \bot \mid A \mid \phi \to \phi \mid \modal[s]\phi,
  \end{gather*}
  where $A \in \mathcal V$, and $s \partof \mathcal S$ is an instance,
  i.e., nonempty.

  We make use of additional logical operators that are defined
  as abbreviations in the usual way, e.g., $\lnot\phi$ abbreviates
  $\phi\to\bot$ and $\modal<s>\phi$ abbreviates $\lnot\modal[s]\lnot\phi$.  
  The logical connective $\to$ is treated as right-associative and
  the operator precedence is defined according the descending sequence
  $(\modal[s], \modal<s>, \lnot, \lor, \land, \to, \liff)$.
\end{definition}

\begin{definition}\label{def:axiomsystem}
  Let $\mathcal S$ be a finite set of atoms and $\mathcal V$ be an
  enumerable set of propositional variables.
  The logic $K_{\mathcal S}^D$ is given by the rules and axioms
  for all nonempty subsets $s$ and $t$ of $\mathcal S$
  \begin{align*}
    &\text{from $\vdash \phi$ and $\vdash \phi\to\psi$ infer
      $\vdash \psi$;}\tag{\axiomdef{MP}}\\
    &\text{from $\vdash \phi$ infer $\vdash \modal[s]\phi$;}\tag{\axiomdef{Nec}}\\
    &\vdash \phi\quad\text{ if $\phi$ is a substitution instance of a propositional tautology;}\tag{\axiomdef{0}}\\
    &\vdash \modal[s](\phi \to \psi) \to \modal[s]\phi \to \modal[s]\psi;
    \tag{\axiomdef{K}}\\
    &\vdash \modal[s]\phi \to \modal[t]\phi\quad\text{ if $s \partof t$.}
    \tag{\axiomdef{Dist}}
  \end{align*}
\end{definition}
Note that $K_{\mathcal S}^D$ is the simplest normal multi-modal logic
in which the principle of information distribution \axiom{Dist'} holds,
and in \cite{fagin2003} it is referred to as $K^D_n$ where $n$
denotes the number of atoms.

\subsection{Semantics of \texorpdfstring{$K^D_{\mathcal S}$}{KDS}}
It is a well-known result (e.g.\
\cite{halpern1985,fagin2003,gerbrandy1998})
of modal logic that $K^D_{\mathcal S}$ is sound, complete, and also
decidable for the following interpretation over certain Kripke structures.
\begin{definition}\label{def:Kripke structure}
  Let $\mathcal S$ be a nonempty set of modalities.
  A \emph{multi-modal Kripke structure for $K^D_{\mathcal S}$} is the
  tuple $M = (\Omega,\mapsto,\pi)$, where
  \begin{enumerate}[(i)]
  \item $\Omega$ is the nonempty set of \emph{possible worlds};
  \item $\mapsto$ is a function that maps any atom $s\in\mathcal S$
    to a binary relation $\mapsto_{s}$ on $\Omega$,
    called the \emph{accessibility relation} of $s$.
    The extension of $\mapsto$ to all instances is
    given by the intersection of their atomic components, i.e.,
    for any instance $s\partof\mathcal S$ we have
    ${\mapsto_s} := \bigcap_{t\in s} {\mapsto_t}$;
  \item $\pi$ is a function that maps any possible world $\omega$ of $\Omega$
    to a subset of $\mathcal V$, $\pi$ is called the \emph{truth assignment}.
  \end{enumerate}

  For any Kripke structure $M$ as above,
  $\omega\in\Omega$, $A\in\mathcal V$, and
  any instance $s\partof \mathcal S$ we define:
  \begin{align*}
    (M,\omega) &\not\models\bot,\\
    (M,\omega) &\models A \iff A\in\pi(\omega),\\
    (M,\omega) &\models \phi\to\psi \iff (M,\omega) \models \phi \text{ implies } (M,\omega) \models \psi,\\
    (M,\omega) &\models \modal[s]\phi \iff
    \text{for all } \omega'\in\Omega \text{ with } \omega \mapsto_s \omega'
    \text{ it holds } (M, \omega') \models \phi.
  \end{align*}
  Whenever the model relation $(M,\omega) \models \phi$ holds, $(M,\omega)$
  is a said to be a \emph{pointed model} of $\phi$. Whenever there exists
  $\omega$ with $(M,\omega) \models \phi$, we say that $\phi$ is
  \emph{satisfiable} in $M$ and write $M\models\phi$ for short.
  We say that $\phi$ is \emph{valid},
  denoted by $\models \phi$, if and only if $M \models \phi$ holds for all
  Kripke structures $M$ as above.
\end{definition}

Let $M$ be a Kripke structure as above.
The definition captures the following properties of information.
An instance $s$ has information $\phi$ at $\omega$, i.e.,
$(M,\omega)\models \modal[s]\phi$, if and only if
$\phi$ holds at all possible worlds that $s$ has access to from $\omega$.
An instance  $s$ considers an information $\phi$ as possible
at $\omega$, i.e., $(M,\omega)\models \modal<s>\phi$,
if only if $s$ has access from $\omega$ to at least one possible world
$\omega'$ where $\phi$ holds. As $\bot$ cannot be assigned to true,
there is no possible world at which $\bot$ holds and $\top$ holds at any
possible world.
It follows that an instance $s$ has inconsistent information at $\omega$,
formally $(M,\omega)\models \modal[s]\bot$, if and only if there is no possible
world that is accessible from $\omega$ for $s$. Now, we specify exactly
what it means when we speak of consistent or inconsistent instances.
An instance $s$ is called \emph{inconsistent} in the model $M$ if and only if
$M \models \modal[s]\bot$ holds. Otherwise, $s$ is called \emph{consistent}
and the dual statement $M\models\modal<s>\top$ holds.

\subsection{Running Example Revisited}
We return to the running example and consider the following derivation
from $\Sigma$ that shows that $K^D_{\mathcal S}$ indeed allows us to identify
those atoms that contribute to a contradiction, i.e.,
there is an instance $s\partof\mathcal S$ such that $\Sigma \vdash \modal[s]\bot$ holds.\footnote{%
The notation $\Sigma \vdash \phi$ means that there exists a finite subset
$\{\sigma_1,\dots,\sigma_m\}\subseteq \Sigma$ such that $\vdash (\sigma_1 \land \dots \land \sigma_m) \to \phi$.}
\begin{align}
  &
  \modal[r_3,g_1] \lnot\texttt{brake}
  &&\text{from $\modal[r_3]$ and $\modal[g_1]$,}
  \label{eq:simple1}\\
  &
  \modal[o_1,r_1](\texttt{brake} \lor \texttt{changeLane})
  &&\text{from $\modal[o_1]$ and $\modal[r_1]$,}
  \label{eq:simple2}\\
  &
  \modal[o_1,r_1,r_3,g_1] \texttt{changeLane}
  &&\text{from \eqref{eq:simple1} and \eqref{eq:simple2},}
  \label{eq:simple3}\\
  &
  \modal[r_4,g_2](\texttt{changeLane} \to \texttt{largeGapAvailable)}
  &&\text{from $\modal[r_4]$ and $\modal[g_2]$,}
  \label{eq:simple4}\\
  &
  \modal[o_1,r_1,r_3, r_4, g_1, g_2] \texttt{largeGapAvailable}
  &&\text{from \eqref{eq:simple3} and \eqref{eq:simple4},}
  \label{eq:simple5}\\
  &
  \modal[o_1, o_2, r_1, r_3, r_4, g_1, g_2] \bot
  &&\text{from $\modal[o_2]$ and \eqref{eq:simple5}.}\label{eq:simple6}
\end{align}
Eventually, in \eqref{eq:simple6} we derived that the instance
$\{o_1, o_2, r_1, r_3, r_4, g_1, g_4\}$ has contradicting information.
We are tempted to say that the entirety of its atoms
$o_1$, $o_2$, $r_1$, $r_3$, $r_4$, $g_1$, $g_4$ provides a justification
for a contradiction.
Let us give in to temptation and look at the intermediate derivation steps.
In line \eqref{eq:simple1} we see that rule $r_3$ and goal $g_1$
justify not to brake, in line \eqref{eq:simple3} we see
that $o_1$, $r_1$, $r_3$, $g_1$ together justify to change lane.
However, line \eqref{eq:simple5} is a bit out of the ordinary as 
we usually would not accept rules and goals as a justification
for an observation. For the present we leave it that way
and take it as an indicator to introduce an additional order
over the atoms later on. 

\subsection{Relation to Justification Logic}
The reference to justifications in the previous section was given with full intent.
Indeed, for each line \eqref{eq:simple1}--\eqref{eq:simple6}
an analogous derivation is also possible in justification logic
\cite{artemov:logic-justification:2008,artemov:jck:2006}
(with axiomatically appropriate constant specification).
To see this, we briefly outline Artemov's justification logics.

Justification logics \cite{artemov:jck:2006} are variants
of epistemic modal logics
where the modal operators of knowledge and belief are unfolded into
justification terms. Hence, justification logics allow a complete
realisation of Plato's characterisation of knowledge as justified true belief.
A typical formula of justification logic has the form $s\::\phi$,
where $s$ is a justification term built from justification constants, and it
is read as ``$\phi$ is justified by $s$''.
The basic justification logic \system{J0} results from extending propositional
logic consisting of \axiom{0} and \axiom{MP} by the application axiom and
the sum axioms
\begin{align*}
  &\vdash s\::(\phi\to\psi)\to t\::\phi\to(s\cdot t)\::\psi, \tag{\axiomdef{Appl}}\\
  &\vdash s\::\phi\to(s + t)\::\phi \text{ and } \vdash s\::\phi\to(t+s)\::\phi,
  \tag{\axiomdef{Sum}}
\end{align*}
where $s$, $t$, $[s\cdot t]$, $[s+t]$, and $[t+s]$ are justification terms
that are assembled from justification constants using the operators $+$
and $\cdot$ according to the axioms.
Justification logics tie the epistemic tradition together
with proof theory. Justification terms are reasonable
abstractions for constructions of proofs. If $s$ is a proof of
$\phi\to\psi$ and $t$ is a proof of $\phi$, then the application
axiom postulates that there is a common proof, namely $s\cdot t$, for $\psi$.
Moreover, if we have a proof $s$ for $\phi$ and some proof $t$, then
the concatenations of both proofs, $s+t$ and $t+s$, are still proofs for $\phi$.
An important predecessor of justification logic is the logic of explicit proofs,
$LP$ \cite{artemov2001explicit}, which from today's perspective turns out to be an extension of $J0$.

The key observation here is that \axiom{Appl} is a variant of the principle of
information distribution \axiom{Dist'}. Hence, many promising features noted so far
immediately carry over to $K^D_{\mathcal S}$. However, there is a subtle difference.
While in $K_{\mathcal S}^D$ grouping of modalities is an idempotent,
commutative, and associative operation, this is not the case for $\cdot$. 
The justification terms $s\cdot t$, $s \cdot (t\cdot t)$, and $t \cdot s$ are
pairwise different. In $K^D_{\mathcal S}$ such a fine grained distinction
is not possible. All justification terms noted before coincide
to the instance $\{s,t\}$ in $K_{\mathcal S}^D$. 

All substitution instances of classical logical tautologies,
like $A \lor \lnot A$ and $s\::A \lor \lnot s\::A$, are provable
in justification logics.
But in contrast to modal logics, justification logics do
not have a necessitation rule. The lack of the necessitation rule allows
justification logics to break the principle of logical awareness, as 
$s\::(A \lor \lnot A)$ is not necessarily provable for an arbitrary
justification term $s$.
Certainly, restricting the principle of logical awareness is attractive
to provide a realistic model of restricted logical resources.
Since we are also interested in revealing and resolving contradictions,
restricting the principle of logical awareness
only would hide latent contradictions. E.g., there is no obvious way
to derive a justification of a contradiction from $s\::(A \land B)$ and
$t\::\lnot A$ without using additional logical resources. Hence, we claim
that the principle of logical awareness is indispensable to discover
all contradictions that are logically possible.

Nevertheless, justification logic can simulate unrestricted logical awareness
by adding proper axiom internalisation rules
$\vdash e\::\phi$ for all axioms $\phi$ and justification constants
$e$. In such systems a weak variant
of the necessitation rule of modal logic holds: for any derivation $\vdash \phi$
there exists a justification term $t$ such that $\vdash t\::\phi$ holds.
Since $\phi$ was derived using axioms and rules only,
also the justification term $t$ is exclusively built from justification constants
dedicated to the involved axioms. Beyond that, $t$ is hardly informative
as it does not help to reveal \emph{extra-logical sources} of a contradiction.

%% file: more_logic.tex
\section{Towards a Logic of Justifiability}\label{sec:justifiability}
Our analysis so far has shown that in $K_{\mathcal S}^D$ instances
play the role of justifications. Or, to pronounce that an instance is the
necessary quality of a justification,
an information \emph{$\phi$ is justifiable} if and only if
\emph{there exists an instance} that is consistent and has the information
$\phi$.
Hence, the canonical next step is
to introduce quantification over instances into our logic.
We introduce a variable modal operators $x$ and the quantifiers
$\forall$ and $\exists$ into our logic and call this extension
$QK^D_{\mathcal S}$. On the justification logic side, a similar approach has been
explored in Fitting's paper \cite{fitting2008quantified},
in which quantification over evidence is introduced for the logic $LP$,
yielding the logic $QLP$. However, to the author's
best knowledge it is still unknown whether $QLP$ is decidable. 

\begin{definition}
  Let $\mathcal V$ be an enumerable set of propositional variables and
  $\mathcal S$ be a finite set of atomic modalities.
  A formula $\phi$ of $QK^D_{\mathcal S}$ is any expression that satisfies the
  Backus-Naur form
  \begin{gather*}
    \phi ::= \bot \mid A \mid \phi \to \phi \mid \modal[s]\phi \mid \modal[x]\phi \mid \forall s \partof x \partof t (\phi) \mid \forall x \partof t (\phi),
  \end{gather*}
  where $A \in \mathcal V$, $s \partof \mathcal S$ and $t\partof \mathcal S$
  are instances, and $x$ is the variable modal operator.
  In formulae of the form $\forall s \partof x \partof t (\phi)$ we call $s$
  the lower bound and $t$ the upper bound.
  The formula $\forall x \partof t (\phi)$
  has only an upper bound. The unbounded quantification in $\forall x (\phi)$
  is short for $\forall x \partof \mathcal S (\phi)$.
  As usual, $\exists$ denotes the dual of $\forall$. 

  The variable $x$ is free in the formula $\phi$ if there is at least one
  occurrence of $x$ in $\phi$ that is not within the scope of a quantifier.
  By $\phi[s/x]$ we denote the formula obtained by replacing all free
  occurrences of the variable modal operator $x$ by the instance $s$.
  A closed formula is a formula where $x$ is not free. 
\end{definition}

Note that the quantification in $QK^D_{\mathcal S}$ is very restrictive, as
there is only one variable symbol. However, together with the fact
that $\mathcal S$ is a finite set this allows for an easy transfer principle:
Any closed formula of $QK^D_{\mathcal S}$ can
be transferred to a formula of $K^D_{\mathcal S}$ by substituting 
$\exists s \partof x\partof t(\phi)$ with the
finite disjunction $\bigvee_{s\partof r \partof t} \phi[r/x]$,
and substituting $\forall s\partof x\partof t(\phi)$ with the finite
conjunction $\bigwedge_{s\partof r \partof t} \phi[r/x]$.
Hence, there is no need to provide an extension of the axiomatisation or the
semantics. The soundness, completeness, and decidability results of
$K^D_{\mathcal S}$ carry over to $QK^D_{\mathcal S}$ directly.
Note that the complexity of a decision procedure might change due to the
exponential blow up of the formula size by the translation principle
given above. However, complexity considerations are beyond the introductory
character of this paper.

In the following section, we will report a prototypical implementation of a
solver accepting formulae of $QK^D_{\mathcal S}$. In particular, it
avoids the naive explicit enumeration of the instances indicated above.
Since the solver searches for a satisfying model of a given formula,
we feed it with the negation of a formula and use that
the unsatisfiability of the negated formula is equivalent to the provability
of the formula under consideration.

\subsection{A Tableau for \texorpdfstring{$QK^D_{\mathcal S}$}{QKDS}}\label{sec:solver}
We use a prefixed tableau prover
\cite{massacci2000single,blackburn2006handbook,sebastiani2009automated}
for $QK^D_{\mathcal S}$ and reduce the modal satisfiability problem to a
Boolean satisfiability problem by using a similar incremental translation
and solving method as has been proposed in \cite{kaminski2013inkresat}.
Due to the usage of prefixes, a tableau proof consists of a single tree, see \cite{fitting1983proof}.
The tableau starts with a single branch that contains the input formula in negation normal form.
The tableau rules are then applied to the branches of the tree until either all branches are
closed, i.e., each branch contains a literal and its negation, or there exists at least one branch
that remains open as all possible rules have been applied to the formulae of the branch.
In the former case, the input formula is unsatisfiable. In the latter case, the
input formula is satisfiable and the open branch provides a partial model of the input formula.

Let $\mathcal S=\{s_1,\dots,s_n\}$ be a set of $n$ atoms and 
$\mathbf p = (\mathbf p_{s_1}, \dots \mathbf p_{s_n})$ a vector of $n$
Boolean variables. Any instance $s\partof\mathcal S$ corresponds
to a particular truth assignment of $\mathbf p$, i.e., $p_{s_i}$
is assigned to true if and only if $s_i\in s$.
We use Boolean formulae over the variables in $\mathbf p$ to constrain
the set of feasible truth assignments. E.g.,
$\bigwedge_{s_i\in s} p_{s_i} \land \bigwedge_{s_i\not\in s} \lnot p_{s_i}$ uniquely determines the assignment
corresponding to $s\partof \mathcal S$ as above, and the formula $\bigvee_{s_i\in\mathcal S} p_i$
encodes a nonempty subset of $\mathcal S$, i.e., an arbitrary instance of $\mathcal S$.

We extend the Boolean satisfiability problem to support several variable assignments.
Each possible world $\omega$ is assigned with an individual Boolean vector $\mathbf p=\mathbf p(\omega)$,
and Boolean formulae over $\mathbf p$ constrain the set of instances that have access to $\omega$.
E.g., consider the instance $s=\{s_1,s_n\}$. The formula $p_{s_1} \land p_{s_n}$ ensures that any feasible
assignment for $\mathbf p$ correspond to a superset of $s$, as at least $p_{s_1}$ and $p_{s_n}$ have to be true under the given constraint.
Formulae of this kind are denoted by $s \partof \mathbf p$ and used to
ensure that at least $s$ has access to $\omega$ (and possibly also some $t$
with $s\partof t$).
It follows that the solver only needs to maintain a single general accessibility
relation $\mapsto$. The individual accessibility $\omega\mapsto_{s}\omega'$
is represented by $\omega \mapsto \omega'$ and $\omega' \models s \partof \mathbf p$.
An immediate consequence is that $\omega\mapsto_{s}\omega'$ is represented
if only if $\omega\mapsto_{s_i}\omega'$ is represented for all atoms $s_i$ of $s$.
Hence, the semantic compatibility of this approach is ensured.

In addition, the solver supports an assignment $\beta$ for
the variable modality $x$.  That is, $\beta$ is an assignment for the
Boolean vector $\mathbf x = \{x_{s_1},\dots,x_{s_n}\}$. E.g., consider
again the instance $s=\{s_1,s_n\}$. The formula
$\lnot x_{s_2} \land \lnot x_{s_3} \land \dots \land \lnot x_{s_{n-1}}$ ensures
that any assignment of $\mathbf x$ corresponds to
a subset of $s$, as only $x_{s_1}$ and $x_{s_n}$ can be true under this
assignment. By $\mathbf x \partof s$ we denote the Boolean formula
as before together with the additional constraint that $\beta$ may not
correspond to the empty set. We use $\mathbf x \partof s$
to encode the upper bound restriction of an existential quantifier.

The prefixes in this tableau consists of pairs $(\omega,\beta)$,
where $\omega$ denotes a possible world and $\beta$ denotes an assignment
for the variable $x$. We write $(\omega,\beta)\models\phi$ to denote
that $(\omega,\beta)$ is the prefix of $\phi$. The tableau makes use
of an explicit representation of the accessibility relation, i.e.,
it maintains a map $\mapsto$ that allows it to look up all possible
worlds $\omega'$ that are reachable from $\omega$. We write $\omega\mapsto\omega'$
to denote that $\omega'$ is reachable from $\omega$.

Note that, for the clarity of presentation, we only give
the rules for upper bounded quantifiers.

\begin{itemize}
\item The $\land$ and $\lor$-rules are classical rules for a Boolean tableau.
\begin{gather*}
  (\,\land\,)\
  \begin{array}{c}
    (\omega,\beta) \models \phi \land \psi\\
    \midrule
    (\omega,\beta) \models \phi\\
    (\omega,\beta) \models \psi
  \end{array}
  \qquad\text{and}\qquad
  (\,\lor\,)\
  \begin{array}{c}
    (\omega,\beta) \models \phi \lor \psi\\
    \midrule
    (\omega,\beta) \models \phi \quad\mid\quad (\omega,\beta) \models \psi
  \end{array}.
\end{gather*}
\item The following $\modal<s>$- and $\modal[s]$-rules are adapted from \cite{kaminski2013inkresat}.
  While the $\modal<s>$-rule is transferred to the tableau directly,
  the $\modal[s]$-rule
  is extended to supports the principle of information distribution. This extension was found
  in \cite[Ch. 2,  pp. 122--123]{blackburn2006handbook}.
  
  The $\modal<s>$-rule introduces a new possible world $\omega'$
  and adds $\omega\mapsto\omega'$ to the accessibility relation.
  The constraint $s\partof \mathbf p$ stipulates that at least
  $s$ has access to $\omega'$.
  The $\modal[s]$-rule takes a formula $\modal[s]\phi$ and an accessible
  world $\omega'$ and adds the prefixed formula
  $(\omega',\beta) \models s\partof \mathbf p \to \phi$
  to the current branch. Hence, whenever $s$ has access to $\omega'$ then
  $\phi$ has to hold. 
\begin{gather*}
  (\,\modal<s>\,)\ 
  \begin{array}{c}
    (\omega,\beta) \models \modal<s>\phi\\
    \midrule
    (\omega',\beta) \models s\partof\mathbf p \land \phi\\
    \omega \mapsto \omega',\quad \text{$\omega'$ new}
  \end{array}
  \qquad\text{and}\qquad
  (\,\modal[s]\,)\
  \begin{array}{c}
    (\omega,\beta) \models \modal[s]\phi,\quad \omega\mapsto\omega'\\
    \midrule
    (\omega',\beta) \models s\partof\mathbf p \to \phi
  \end{array}.
\end{gather*}
\item The following $\modal<x>$- and $\modal[x]$-rules for variable modal
  operators are similar to their variants for constant modal operators but
  depend on
  the assignment $\beta$ for $\mathbf x$.
  Readers may convince themselves that the constraint
  $\mathbf x \partof \mathbf p$
  can actually be encoded into a Boolean satisfiability solver using the 
  encoding scheme as indicated above.
\begin{gather*}
  (\,\modal<x>\,)\ 
  \begin{array}{c}
    (\omega,\beta) \models \modal<x>\phi\\
    \midrule
    (\omega',\beta) \models \mathbf x\partof\mathbf p \land \phi\\
    \omega \mapsto \omega',\quad \text{$\omega'$ new}
  \end{array}
  \qquad\text{and}\qquad
  (\,\modal[x]\,)\
  \begin{array}{c}
    (\omega,\beta) \models \modal[x]\phi,\quad \omega\mapsto\omega'\\
    \midrule
    (\omega',\beta) \models \mathbf x\partof\mathbf p \to \phi
  \end{array}.
\end{gather*}
\item The $\exists$-rule introduces a new valuation $\beta'$ for $\mathbf x$
  and puts the constraint $\mathbf x \partof s$ on this valuation.
  The $\forall_0$-rule generates a substitution instance where the variable
  is substituted by its upper bound.
\begin{gather*}
  (\,\exists\,)\
  \begin{array}{c}
    (\omega,\beta) \models \exists x \partof s (\phi)\\
    \midrule
    (\omega,\beta') \models \mathbf x \partof s \land \phi\\
    \text{$\beta'$ new}
  \end{array}
  \qquad\text{and}\qquad
  (\,\forall_0\,)\
  \begin{array}{c}
    (\omega,\beta) \models \forall x \partof s (\phi)\\
    \midrule
    (\omega,\beta) \models \phi[s/x]\\
  \end{array}.
\end{gather*}
Finally, the $\forall_s$-rule generates a constrained substitution instance
$\phi[t/x]$ whenever there occurs another formula of the form $\modal<t>\psi$ or $\modal[t]\psi$ that has the same $\omega$-prefix in the tableau. The constraint
$t\partof s$ ensures that $\phi[t/x]$ has to hold at least if $t \partof s$.
For the implementation, this constraint can be checked before
application of the rule. 
Similarly, the $\forall_x$-rule takes the valuation $\beta'$ of any variable modal operator
that has the same $\omega$-prefix and adds the constrained formula $\mathbf x \partof s \to \phi$. That is, if $\beta'$ assigns $x$
to some subinstance $t\partof s$, then $\phi$ has to hold, and
any free occurrence of $x$ in $\phi$ is now interpreted as $t$ by $\beta'$.
\begin{gather*}
  (\,\forall_s\,)\
  \begin{array}{c}
    (\omega,\beta) \models \forall x \partof s (\phi),\quad
    (\omega,\beta') \models \modal<t>\psi \text{ or }
    (\omega,\beta') \models \modal[t]\psi\\
    \midrule
    (\omega,\beta') \models t\partof s \to \phi[t/x]
  \end{array}
  \qquad\text{and}
  \\
  \\
  (\,\forall_x\,)\
  \begin{array}{c}
    (\omega,\beta) \models \forall x \partof s (\phi),\quad
    (\omega,\beta') \models \modal<x>\psi \text{ or }
    (\omega,\beta') \models \modal[x]\psi\\
    \midrule
    (\omega,\beta') \models \mathbf x \partof s \to \phi\\
  \end{array}.
\end{gather*}
\end{itemize}

It remains to argue that this tableau is a sound and complete decision procedure
for $QK^D_{\mathcal S}$. We use that the rules for $\land$, $\lor$, $\modal<s>$ and $\modal[s]$ provide a decision procedure for $K^D_{\mathcal S}$, see \cite[Ch. 2]{blackburn2006handbook}.
While the existential fragment of this logic is quite directly translated
into a Boolean satisfiability problem and not further discussed here, we show
that the rules for the universal quantifier are complete for $QK^D_{\mathcal S}$.
Clearly, a $\forall$-rule that produced all possible substitution
instances would yield a complete decision procedure. However,
the $\forall_0$-rule ensures that at least the substitution instance
of the upper bound is produced.
All other substitution instances are produced only when a corresponding
instance occurs in the branch with the same $\omega$-prefix.
This approach is sufficient. Assume we found
a pointed model $(M,\omega)$ for some formula $\phi$ where $s$ is
a least instance occurring in $\phi$. Then for any instances $t\partof s$
we can substitute $t$ for $s$ in $\phi$ and construct
a pointed model $(M',\omega)$ for $\phi[t/s]$ by
setting $\omega \mapsto_t \omega'$ in the new model
if and only $\omega \mapsto_s \omega'$ in the original model.
But this means that either there is a specific instances that contradicts
a bounded $\forall$-statement or it is sufficient to show the satisfiability of the
substitution instance for the upper bounding instance.

The solver is implemented in C++ and uses minisat to solve the Boolean satisfiability problem. An early prototype is available at \url{https://vhome.offis.de/~willemh/episat/}. The site will be updated regularly.

\subsection{Justifiability in Practice} 
Let us use quantified expression to analyse our running example with the help
of the solver. Note that these examples are also distributed with the prototype.
Due to nondeterminism the concrete partial model may change in the satisfiable case.

\begin{enumerate}
\item The formula $\exists x (\modal[x]\bot)$ holds if and only
  if there is some instance that has inconsistent information.
  As we already have seen, the automated driving system of the vehicle
  has inconsistent information and we can expect that
  $\Sigma \to \exists x (\modal[x]\bot)$ is derivable.
  Indeed, as we already have shown above, it suffices to choose
  $x = [o_1, o_2, r_1, r_3, r_4, g_1, g_2]$ to see that
  the existential formula holds. We use the solver and
  verify that the negation
  $\Sigma \land \forall x (\modal<x>\top)$ is unsatisfiable.
  Indeed, the solver cannot find a model and returns the unsatisfiability result.
  Notice the difference between the derivation in a proof calculus
  as given above and the satisfiability solver.
  While we could directly read off the instance having
  inconsistent information from the derivation,
  the solver provides us no hint in case of unsatisfiability.
\item Is there at least some instance that has consistent information?
  I.e., can we satisfy
  $\Sigma \land \exists x (\modal<x>\top)$?\\
  In this case, the solver returns a satisfying model consisting of
  two possible worlds $\omega_0$ and $\omega_1$, where $\omega_1$
  is accessible for $o_1$ from $\omega_0$. Further, it returns the
  partial truth assignment
  $\omega_1\models \{\lnot\texttt{beConsiderate},
  \lnot\texttt{beFast}, \texttt{laneBlocked}\}$ and the valuation $x=o_1$.
  This is indeed a model for the formula, as there is only one world
  $\omega_1$ accessible from the actual world $\omega_0$, and only $o_1$ can
  access it. Hence, $\modal<o_1>\top$ holds. Moreover, also the subformula 
  $\modal[o_1]\texttt{laneBlocked}$ of $\Sigma$ is satisfied, as
  $\texttt{laneBlocked}$ holds in any $o_1$-accessible possible world which 
  precisely is $\omega_1$. All other subformulae of $\Sigma$
  are trivially satisfied by the semantics of the modal operator
  $\modal[.]$. As this is an obvious but somewhat uninformative result, we turn
  to the next analysis.
\item Do the rules have inconsistent information?
  Can we satisfy
  $\Sigma \land \forall x \partof \{\mathlist{r_1,r_2,r_3,r_4}\} (\modal<x>\top)$?\\
  The solver returns a satisfying model that consists of
  two possible worlds $\omega_0$ and $\omega_1$ with
  $\omega_0 \mapsto_{\{r_1,r_2,r_3,r_4\}} \omega_1$ and the
  partial truth assignment
  $\omega_1\models \{\mathlist{
    \lnot\texttt{changeLane},
    \lnot\texttt{largeGapAvailable},
    \lnot\texttt{laneBlocked},
    \lnot\texttt{brake},
    \lnot\texttt{beFast},
    \lnot\texttt{beConsiderate},
    \lnot\texttt{gapAvailable}
  }\}$.
  To convince ourselves that this is again a model, we have to take
  the model properties into account: As $\omega_1$ is
  accessible for $\{r_1,r_2,r_3,r_4\}$, it is also accessible for all instances
  $s \partof \{r_1,r_2,r_3,r_4\}$.
\item The previous inquiries were mainly motivated to give the reader
  an overview of the possibilities of the solver and enable him to interpret
  the output models. Let us now turn to a more interesting inquiry that
  analyses the role of the user specified goals.
  Can we find a model where an instance exists that has consistent information
  and includes all rules and observations? Can we satisfy $\Sigma \land
  \exists  \{r_1,r_2,r_3,r_4,o_1,o_2\} \partof x \partof \mathcal S(\modal<x>\top)$?\\
  The solver returns the model
  $\omega_0 \mapsto_{\{r_1,r_2,r_3,r_4,o_1,o_2\}} \omega_1$ with the
  truth assignment $\omega_1 \models \{\mathlist{
    \texttt{changeLane},
    \texttt{laneBlocked},\hfill
    \lnot\texttt{largeGapAvailable},\hfill
    \lnot\texttt{beConsiderate},\hfill
    \texttt{gapAvailable},\hfill
    \lnot\texttt{beFast},\hfill
    \lnot\texttt{brake}
  }\}$
  and the valuation
  $x= \{\mathlist{r_1,r_2,r_3,r_4,o_1,o_2}\}$.
  This is a quite convincing result. If the automated driving system would
  have given up the user specified goals, it could have found a solution
  where it restrained from braking and changed the lane.
\item We ask whether we can find an instance $x$ that includes the rules,
  has consistent information and considers braking a possible manoeuvre.
  Is it possbile to satisfy
  $\Sigma \land \exists \{r_1,r_2,r_3,r_4\} \partof x \partof \mathcal S\linebreak
  (\modal<x>\texttt{brake} \land \modal[x](\top))$?\\
  The solver returns the model
  $\omega_0 \mapsto_{\{r_1,r_2,r_3,r_4,o_1,o_2,g_2\}} \omega_1$ with the
  truth assignment $\omega_1 \models \{\mathlist{
    \lnot\texttt{beFast},
    \texttt{beConsiderate},\hfill
    \lnot\texttt{changeLane},\hfill
    \texttt{laneBlocked},\hfill
    \texttt{gapAvailable},\hfill
    \lnot\texttt{largeGapAvailable},\hfill
    \texttt{brake}}\}$\linebreak
  and $x=\{r_1,r_2,r_3,r_4,o_1,o_2,g_2\}$.
  That is, roughly speaking, if braking would be possible,
  than $\{r_1,r_2,r_3,r_4,o_1,o_2,g_2\}$ can be made a consistent instance
  with the information as given for $\omega_1$.
\end{enumerate}

%% file: conclusion.tex
\section{Conclusion}\label{sec:conclusion}
I presented the logic $QK^D_{\mathcal S}$ that has strong connections
to justification logic. In contrast to justification logic,
$QK^D_{\mathcal S}$ supports for the quantification over a variable modality.
In combination with the proposed tableau-based satisfiability solver,
$QK^D_{\mathcal S}$ can be used to identify information components justifying
given statements. Moreover, I reported on a prototypical implementation of
the satisfiability solver accompanied by some exemplary inquiries along
our running example.

An extensive comparison with other solvers remains as future work.
To the best of the author's knowledge this is the
first proposal of a decidable theory supporting quantification over
modalities. Certainly, there is
a lack of comparable solvers and benchmarks. An obvious approach
would be to use the translation of bounded quantification to 
finite dis- and conjunctions as indicated above and compare 
against solvers for $K^D_n$. Unfortunately,
such solvers are rare. Neither InKreSAT \cite{kaminski2013inkresat}, nor
solvers for description logic like Racer \cite{HHMW12} or
FaCT++ \cite{10.1007/11814771_26}
support the principle of information distribution, as the intersection of
roles is not part of the description logics $\mathcal{SHOIQ}$ or $\mathcal{SRIQ}(\mathcal D)$.
The only promising candidate known to the author is the first-order logic theorem prover
SPASS that fully supports Boolean modal logic of which $K^D_n$ is a fragment
\cite{10.1007/978-3-540-73595-3_38}.
As the tableau for the multi-modal fragment $K_n$ without the
principle of information distribution follows the same rules as InKreSAT,
a detailed comparison of this fragment has been postponed so far,
as we do not expect any interesting insights of such a comparison
beside of potentially revealing weak points of the prototypical implementation.

Another important aspect is to investigate the theoretical complexity of
the satisfiability problem for $QK^D_{\mathcal S}$ 
which is at least PSPACE-hard \cite{fagin2003}, and the
complexity of the presented tableau calculus.

The current focus of development is mainly on the extension of this logic.
The prototype of the solver already supports a partial order of the atoms in
$\mathcal S$. Such a partial order implies that the set of instances has
the interesting structure of a semilattice presentation
\cite{binczak2007poset,silva_1993,silva2000unique}.
That is, any instance is either irreducible or
it can be written as a finite join of entities. However, the semilattice
may contain entities that are irreducible but not atomic. Nonatomic entities
are of special interest. They inherit all information of smaller instances
but allow for the emergence of novel information, e.g., by quantifying
over their instances. This gives rise to powerful nonmonotonic capability
as entities can mimic reasoning by default, autoepistemic reasoning or
counterfactual reasoning \cite{lewis1973counterfactuals}.
More details are planned for a follow-up paper,
until then readers may convince themselves that the partial order of the atoms
can simply be added to the Boolean formulae encoding instances.

Any expectation that $QK^D_{\mathcal S}$ could be a logic that
resolves inconsistencies fully automatically has to remain unfulfilled.
While this logic provides a rigorous framework in which information atoms
can be held and maintained independently, it cannot decide 
to which atoms should be given preference over others.
Nevertheless, $QK^D_{\mathcal S}$ gives us raise to
a weak notion of belief. We may say that a proposition $\phi$ is believed
in a system $\mathcal S$ if and only if $\phi$ is justifiable and $\lnot\phi$
is not justifiable, i.e., if and only if $\exists x(\modal<x>\top \land \modal[x]\phi) \land \forall x(\modal<x>\top \to \modal<x>\phi)$ holds. To put this
definition in legal terms, $\phi$ is believed if and only if
there is a noncontradictory group of witnesses for $\phi$,
and all other groups of witnesses
either contradict themselves or have no argument against $\phi$.
This concept of belief thus is a reasonable selection of appropriate
justifications, and remains consistent in itself.
A further observation
is that this kind of belief is no longer closed under logical consequence and,
hence, is not a normal modality.

Another important extension in this regard is to allow positive or negative
introspection axioms \cite{fagin2003}
to $QK^D_{\mathcal S}$, i.e., for any $\phi$ and $s\in\mathcal S$
\begin{align}
  &\vdash \modal \modal[s]\phi \to \modal[s]\modal[s]\phi \text{ and} \tag{\axiomdef{4}}\\
  &\vdash \modal<s>\phi \to \modal[s]\modal<s>\phi. \tag{\axiomdef{5}}
\end{align}
This would yield the logic $QK^D_{\mathcal S}45$ of instances with
additional epistemic power. In contrast, we are less interested in
the extensions
\begin{align}
  &\vdash \modal \modal<s>\top \text{ or }\tag{\axiomdef{D}}\\
  &\vdash \modal[s]\phi \to \phi, \tag{\axiomdef{T}}
\end{align}
as we consider these axioms as too powerful. Logics where
at least one of these principles holds do not allow for contradicting instances
and are therefore of lesser interest for the analysis of justifiability.

A temporal expansion to better understand the dynamic aspects of
information distribution within systems would be very attractive.
The interaction of information and time is certainly an appealing
epistemic challenge here.

A profound philosophical reflection of the proposed approach would require
an intensive examination of Hempel and Oppenheim's concept of
a DN-explanation \cite{hempel1948studies}, Toulmin's concept of an argument
\cite{toulmin}, and also the fallibilist position of Popper's critical rationalism \cite{popper2005logic}, just to name a few.

\paragraph*{Acknowledgements}
I would like to thank the anonymous referees for their valuable comments
and carefully proofreading my work.
I hope that their common desire for better motivation and reflection
of related work has been met.

%% file: main.bbl
\begin{thebibliography}{10}
\providecommand{\bibitemdeclare}[2]{}
\providecommand{\surnamestart}{}
\providecommand{\surnameend}{}
\providecommand{\urlprefix}{Available at }
\providecommand{\url}[1]{\texttt{#1}}
\providecommand{\href}[2]{\texttt{#2}}
\providecommand{\urlalt}[2]{\href{#1}{#2}}
\providecommand{\doi}[1]{doi:\urlalt{http://dx.doi.org/#1}{#1}}
\providecommand{\eprint}[1]{arXiv:\urlalt{https://arxiv.org/abs/#1}{#1}}
\providecommand{\bibinfo}[2]{#2}

\bibitemdeclare{article}{Alur1995}
\bibitem{Alur1995}
\bibinfo{author}{Rajeev \surnamestart Alur\surnameend}, \bibinfo{author}{Costas
  \surnamestart Courcoubetis\surnameend}, \bibinfo{author}{Nicolas
  \surnamestart Halbwachs\surnameend}, \bibinfo{author}{Thomas~A. \surnamestart
  Henzinger\surnameend}, \bibinfo{author}{Pei-Hsin \surnamestart
  Ho\surnameend}, \bibinfo{author}{Xavier \surnamestart Nicollin\surnameend},
  \bibinfo{author}{Alfredo \surnamestart Olivero\surnameend},
  \bibinfo{author}{Joseph \surnamestart Sifakis\surnameend} \&
  \bibinfo{author}{Sergio \surnamestart Yovine\surnameend}
  (\bibinfo{year}{1995}): \emph{\bibinfo{title}{The algorithmic analysis of
  hybrid systems}}.
\newblock {\sl \bibinfo{journal}{Theoretical Computer Science}}
  \bibinfo{volume}{138}(\bibinfo{number}{1}), pp. \bibinfo{pages}{3--34},
  \doi{10.1016/0304-3975(94)00202-T}.

\bibitemdeclare{article}{jsan10030041}
\bibitem{jsan10030041}
\bibinfo{author}{Gleifer~Vaz \surnamestart Alves\surnameend},
  \bibinfo{author}{Louise \surnamestart Dennis\surnameend} \&
  \bibinfo{author}{Michael \surnamestart Fisher\surnameend}
  (\bibinfo{year}{2021}): \emph{\bibinfo{title}{A Double-Level Model Checking
  Approach for an Agent-Based Autonomous Vehicle and Road Junction
  Regulations}}.
\newblock {\sl \bibinfo{journal}{Journal of Sensor and Actuator Networks}}
  \bibinfo{volume}{10}(\bibinfo{number}{3}), \doi{10.3390/jsan10030041}.

\bibitemdeclare{article}{artemov2001explicit}
\bibitem{artemov2001explicit}
\bibinfo{author}{Sergei~N \surnamestart Artemov\surnameend}
  (\bibinfo{year}{2001}): \emph{\bibinfo{title}{Explicit provability and
  constructive semantics}}.
\newblock {\sl \bibinfo{journal}{Bulletin of Symbolic logic}}
  \bibinfo{volume}{7}(\bibinfo{number}{1}), pp. \bibinfo{pages}{1--36},
  \doi{10.2307/2687821}.

\bibitemdeclare{article}{artemov:jck:2006}
\bibitem{artemov:jck:2006}
\bibinfo{author}{Sergei~N. \surnamestart Artemov\surnameend}
  (\bibinfo{year}{2006}): \emph{\bibinfo{title}{Justified common knowledge}}.
\newblock {\sl \bibinfo{journal}{Theoretical Computer Science}}
  \bibinfo{volume}{357}(\bibinfo{number}{1-3}), pp. \bibinfo{pages}{4--22},
  \doi{10.1016/j.tcs.2006.03.009}.

\bibitemdeclare{article}{artemov:logic-justification:2008}
\bibitem{artemov:logic-justification:2008}
\bibinfo{author}{Sergei~N. \surnamestart Artemov\surnameend}
  (\bibinfo{year}{2008}): \emph{\bibinfo{title}{The Logic of Justification}}.
\newblock {\sl \bibinfo{journal}{The Review of Symbolic Logic}}
  \bibinfo{volume}{1}(\bibinfo{number}{4}), pp. \bibinfo{pages}{477--513},
  \doi{10.1017/S1755020308090060}.

\bibitemdeclare{article}{baier2021verification}
\bibitem{baier2021verification}
\bibinfo{author}{Christel \surnamestart Baier\surnameend},
  \bibinfo{author}{Clemens \surnamestart Dubslaff\surnameend},
  \bibinfo{author}{Florian \surnamestart Funke\surnameend},
  \bibinfo{author}{Simon \surnamestart Jantsch\surnameend},
  \bibinfo{author}{Rupak \surnamestart Majumdar\surnameend},
  \bibinfo{author}{Jakob \surnamestart Piribauer\surnameend} \&
  \bibinfo{author}{Robin \surnamestart Ziemek\surnameend}
  (\bibinfo{year}{2021}): \emph{\bibinfo{title}{From Verification to
  Causality-based Explications}}.
\newblock {\sl \bibinfo{journal}{arXiv preprint arXiv:2105.09533}}.

\bibitemdeclare{book}{barwise1989situation}
\bibitem{barwise1989situation}
\bibinfo{author}{Jon \surnamestart Barwise\surnameend} (\bibinfo{year}{1989}):
  \emph{\bibinfo{title}{The situation in logic}}.
\newblock {\sl \bibinfo{series}{CSLI Lecture notes}}~\bibinfo{volume}{17},
  \bibinfo{publisher}{Center for the Study of Language and Information (CSLI),
  Stanford}.

\bibitemdeclare{article}{binczak2007poset}
\bibitem{binczak2007poset}
\bibinfo{author}{G.~\surnamestart Bińczak\surnameend}, \bibinfo{author}{A.B.
  \surnamestart Romanowska\surnameend} \& \bibinfo{author}{J.D.H. \surnamestart
  Smith\surnameend} (\bibinfo{year}{2007}): \emph{\bibinfo{title}{Poset
  extensions, convex sets, and semilattice presentations}}.
\newblock {\sl \bibinfo{journal}{Discrete Mathematics}}
  \bibinfo{volume}{307}(\bibinfo{number}{1}), pp. \bibinfo{pages}{1--11},
  \doi{10.1016/j.disc.2006.09.021}.

\bibitemdeclare{book}{blackburn2006handbook}
\bibitem{blackburn2006handbook}
\bibinfo{author}{Patrick \surnamestart Blackburn\surnameend},
  \bibinfo{author}{Johan \surnamestart van Benthem\surnameend} \&
  \bibinfo{author}{Frank \surnamestart Wolter\surnameend}
  (\bibinfo{year}{2006}): \emph{\bibinfo{title}{Handbook of modal logic}}.
\newblock \bibinfo{publisher}{Elsevier}.

\bibitemdeclare{book}{clarke2018model}
\bibitem{clarke2018model}
\bibinfo{author}{Edmund~M. \surnamestart Clarke~Jr.\surnameend},
  \bibinfo{author}{Orna \surnamestart Grumberg\surnameend},
  \bibinfo{author}{Daniel \surnamestart Kroening\surnameend},
  \bibinfo{author}{Doron \surnamestart Peled\surnameend} \&
  \bibinfo{author}{Helmut \surnamestart Veith\surnameend}
  (\bibinfo{year}{2018}): \emph{\bibinfo{title}{Model checking}},
  \bibinfo{edition}{2nd} edition.
\newblock \bibinfo{series}{Cyper Physical Systems Series},
  \bibinfo{publisher}{MIT press}.

\bibitemdeclare{article}{Damm2012}
\bibitem{Damm2012}
\bibinfo{author}{Werner \surnamestart Damm\surnameend},
  \bibinfo{author}{Henning \surnamestart Dierks\surnameend},
  \bibinfo{author}{Stefan \surnamestart Disch\surnameend},
  \bibinfo{author}{Willem \surnamestart Hagemann\surnameend},
  \bibinfo{author}{Florian \surnamestart Pigorsch\surnameend},
  \bibinfo{author}{Christoph \surnamestart Scholl\surnameend},
  \bibinfo{author}{Uwe \surnamestart Waldmann\surnameend} \&
  \bibinfo{author}{Boris \surnamestart Wirtz\surnameend}
  (\bibinfo{year}{2012}): \emph{\bibinfo{title}{Exact and fully symbolic
  verification of linear hybrid automata with large discrete state spaces}}.
\newblock {\sl \bibinfo{journal}{Science of Computer Programming}}
  \bibinfo{volume}{77}(\bibinfo{number}{10-11}), pp.
  \bibinfo{pages}{1122--1150}, \doi{10.1016/j.scico.2011.07.006}.

\bibitemdeclare{misc}{XAI}
\bibitem{XAI}
\bibinfo{author}{\surnamestart DARPA\surnameend}:
  \emph{\bibinfo{title}{Explainable Artificial Intelligence (XAI) Program}}.
\newblock
  \urlprefix\url{https://www.darpa.mil/program/explainable-artificial-intelligence}.
\newblock \bibinfo{note}{Last accessed on 06/10/2021}.

\bibitemdeclare{inproceedings}{dissing2020implementing}
\bibitem{dissing2020implementing}
\bibinfo{author}{Lasse \surnamestart Dissing\surnameend} \&
  \bibinfo{author}{Thomas \surnamestart Bolander\surnameend}
  (\bibinfo{year}{2020}): \emph{\bibinfo{title}{Implementing Theory of Mind on
  a Robot Using Dynamic Epistemic Logic}}.
\newblock In \bibinfo{editor}{Christian \surnamestart Bessiere\surnameend},
  editor: {\sl \bibinfo{booktitle}{Proceedings of the Twenty-Ninth
  International Joint Conference on Artificial Intelligence, {IJCAI-20}}},
  \bibinfo{publisher}{International Joint Conferences on Artificial
  Intelligence Organization}, pp. \bibinfo{pages}{1615--1621},
  \doi{10.24963/ijcai.2020/224}.

\bibitemdeclare{book}{fagin2003}
\bibitem{fagin2003}
\bibinfo{author}{Ronald \surnamestart Fagin\surnameend},
  \bibinfo{author}{Joseph~Y. \surnamestart Halpern\surnameend},
  \bibinfo{author}{Yoram \surnamestart Moses\surnameend} \&
  \bibinfo{author}{Moshe~Y. \surnamestart Vardi\surnameend}
  (\bibinfo{year}{2003}): \emph{\bibinfo{title}{Reasoning About Knowledge}}.
\newblock \bibinfo{publisher}{MIT Press}, \bibinfo{address}{Cambridge, MA,
  USA}.

\bibitemdeclare{article}{fisher2013verifying}
\bibitem{fisher2013verifying}
\bibinfo{author}{Michael \surnamestart Fisher\surnameend},
  \bibinfo{author}{Louise \surnamestart Dennis\surnameend} \&
  \bibinfo{author}{Matt \surnamestart Webster\surnameend}
  (\bibinfo{year}{2013}): \emph{\bibinfo{title}{Verifying autonomous systems}}.
\newblock {\sl \bibinfo{journal}{Communications of the ACM}}
  \bibinfo{volume}{56}(\bibinfo{number}{9}), pp. \bibinfo{pages}{84--93},
  \doi{10.1145/2494558}.

\bibitemdeclare{book}{fitting1983proof}
\bibitem{fitting1983proof}
\bibinfo{author}{Melvin \surnamestart Fitting\surnameend}
  (\bibinfo{year}{1983}): \emph{\bibinfo{title}{Proof methods for modal and
  intuitionistic logics}}.
\newblock {\sl \bibinfo{series}{Synthese Library}} \bibinfo{volume}{169},
  \bibinfo{publisher}{Springer Science \& Business Media},
  \doi{10.1007/978-94-017-2794-5}.

\bibitemdeclare{article}{fitting2008quantified}
\bibitem{fitting2008quantified}
\bibinfo{author}{Melvin \surnamestart Fitting\surnameend}
  (\bibinfo{year}{2008}): \emph{\bibinfo{title}{A quantified logic of
  evidence}}.
\newblock {\sl \bibinfo{journal}{Annals of Pure and Applied Logic}}
  \bibinfo{volume}{152}(\bibinfo{number}{1}), pp. \bibinfo{pages}{67--83},
  \doi{10.1016/j.apal.2007.11.003}.

\bibitemdeclare{incollection}{Frehse2011}
\bibitem{Frehse2011}
\bibinfo{author}{Goran \surnamestart Frehse\surnameend}, \bibinfo{author}{Colas
  \surnamestart Le~Guernic\surnameend}, \bibinfo{author}{Alexandre
  \surnamestart Donz\'{e}\surnameend}, \bibinfo{author}{Scott \surnamestart
  Cotton\surnameend}, \bibinfo{author}{Rajarshi \surnamestart Ray\surnameend},
  \bibinfo{author}{Olivier \surnamestart Lebeltel\surnameend},
  \bibinfo{author}{Rodolfo \surnamestart Ripado\surnameend},
  \bibinfo{author}{Antoine \surnamestart Girard\surnameend},
  \bibinfo{author}{Thao \surnamestart Dang\surnameend} \& \bibinfo{author}{Oded
  \surnamestart Maler\surnameend} (\bibinfo{year}{2011}):
  \emph{\bibinfo{title}{Space{E}x: Scalable Verification of Hybrid Systems}}.
\newblock In \bibinfo{editor}{Ganesh \surnamestart Gopalakrishnan\surnameend}
  \& \bibinfo{editor}{Shaz \surnamestart Qadeer\surnameend}, editors: {\sl
  \bibinfo{booktitle}{Computer Aided Verification}}, {\sl
  \bibinfo{series}{Lecture Notes in Computer Science}} \bibinfo{volume}{6806},
  \bibinfo{publisher}{Springer, Heidelberg}, pp. \bibinfo{pages}{379--395},
  \doi{10.1007/978-3-642-22110-1\_30}.

\bibitemdeclare{inproceedings}{gerbrandy1998}
\bibitem{gerbrandy1998}
\bibinfo{author}{Jelle \surnamestart Gerbrandy\surnameend}
  (\bibinfo{year}{1998}): \emph{\bibinfo{title}{Distributed knowledge}}.
\newblock In: {\sl \bibinfo{booktitle}{Twendial 1998: Formal Semantics and
  Pragmatics of Dialogue}}, \bibinfo{volume}{98}, pp.
  \bibinfo{pages}{111--124}.

\bibitemdeclare{article}{HHMW12}
\bibitem{HHMW12}
\bibinfo{author}{Volker \surnamestart Haarslev\surnameend},
  \bibinfo{author}{Kay \surnamestart Hidde\surnameend}, \bibinfo{author}{Ralf
  \surnamestart M{\"o}ller\surnameend} \& \bibinfo{author}{Michael
  \surnamestart Wessel\surnameend} (\bibinfo{year}{2012}):
  \emph{\bibinfo{title}{The RacerPro knowledge representation and reasoning
  system}}.
\newblock {\sl \bibinfo{journal}{Semantic Web Journal}}
  \bibinfo{volume}{3}(\bibinfo{number}{3}), pp. \bibinfo{pages}{267--277},
  \doi{10.3233/SW-2011-0032}.

\bibitemdeclare{inproceedings}{halpern1985}
\bibitem{halpern1985}
\bibinfo{author}{Joseph~Y. \surnamestart Halpern\surnameend} \&
  \bibinfo{author}{Yoram \surnamestart Moses\surnameend}
  (\bibinfo{year}{1985}): \emph{\bibinfo{title}{A guide to the modal logics of
  knowledge and belief: preliminary draft}}.
\newblock In: {\sl \bibinfo{booktitle}{Proceedings of the 9th international
  joint conference on Artificial Intelligence -- Volume 1}},
  \bibinfo{publisher}{Morgan Kaufmann Publishers Inc.}, pp.
  \bibinfo{pages}{480--490}.

\bibitemdeclare{article}{hempel1948studies}
\bibitem{hempel1948studies}
\bibinfo{author}{Carl~G. \surnamestart Hempel\surnameend} \&
  \bibinfo{author}{Paul \surnamestart Oppenheim\surnameend}
  (\bibinfo{year}{1948}): \emph{\bibinfo{title}{Studies in the Logic of
  Explanation}}.
\newblock {\sl \bibinfo{journal}{Philosophy of science}}
  \bibinfo{volume}{15}(\bibinfo{number}{2}), pp. \bibinfo{pages}{135--175},
  \doi{10.1086/286983}.

\bibitemdeclare{inproceedings}{kaminski2013inkresat}
\bibitem{kaminski2013inkresat}
\bibinfo{author}{Mark \surnamestart Kaminski\surnameend} \&
  \bibinfo{author}{Tobias \surnamestart Tebbi\surnameend}
  (\bibinfo{year}{2013}): \emph{\bibinfo{title}{InKreSAT: modal reasoning via
  incremental reduction to SAT}}.
\newblock In \bibinfo{editor}{Maria~Paola \surnamestart Bonacina\surnameend},
  editor: {\sl \bibinfo{booktitle}{Automated Deduction -- CADE-24}},
  \bibinfo{organization}{Springer}, pp. \bibinfo{pages}{436--442},
  \doi{10.1007/978-3-642-38574-2\_31}.

\bibitemdeclare{book}{lewis1973counterfactuals}
\bibitem{lewis1973counterfactuals}
\bibinfo{author}{David \surnamestart Lewis\surnameend} (\bibinfo{year}{1973}):
  \emph{\bibinfo{title}{Counterfactuals}}, \bibinfo{edition}{reissued 2001}
  edition.
\newblock \bibinfo{publisher}{Blackwell}.

\bibitemdeclare{article}{massacci2000single}
\bibitem{massacci2000single}
\bibinfo{author}{Fabio \surnamestart Massacci\surnameend}
  (\bibinfo{year}{2000}): \emph{\bibinfo{title}{Single step tableaux for modal
  logics}}.
\newblock {\sl \bibinfo{journal}{Journal of Automated Reasoning}}
  \bibinfo{volume}{24}(\bibinfo{number}{3}), pp. \bibinfo{pages}{319--364},
  \doi{10.1023/A:1006155811656}.

\bibitemdeclare{book}{sae}
\bibitem{sae}
\bibinfo{author}{\surnamestart {On-Road Automated Driving (ORAD)
  committee}\surnameend}: \emph{\bibinfo{title}{Taxonomy and Definitions for
  Terms Related to Driving Automation Systems for On-Road Motor Vehicles}},
  \bibinfo{edition}{2021-04-30} edition.
\newblock \bibinfo{publisher}{SAE International}, \doi{10.4271/J3016\_202104}.

\bibitemdeclare{misc}{CPEC}
\bibitem{CPEC}
\bibinfo{author}{Foundations \surnamestart of~Perspicuous
  Software~Systems\surnameend}: \emph{\bibinfo{title}{Center for Perspicuous
  Computing\,---\,Research}}.
\newblock \urlprefix\url{https://www.perspicuous-computing.science/research/}.
\newblock \bibinfo{note}{Last accessed on 06/10/2021}.

\bibitemdeclare{inproceedings}{platzer2019logical}
\bibitem{platzer2019logical}
\bibinfo{author}{Andr{\'e} \surnamestart Platzer\surnameend}
  (\bibinfo{year}{2019}): \emph{\bibinfo{title}{The logical path to autonomous
  cyber-physical systems}}.
\newblock In: {\sl \bibinfo{booktitle}{International Conference on Quantitative
  Evaluation of Systems}}, \bibinfo{organization}{Springer}, pp.
  \bibinfo{pages}{25--33}, \doi{10.1007/978-3-030-30281-8\_2}.

\bibitemdeclare{book}{popper2005logic}
\bibitem{popper2005logic}
\bibinfo{author}{Karl \surnamestart Popper\surnameend} (\bibinfo{year}{2002}):
  \emph{\bibinfo{title}{The logic of scientific discovery}},
  \bibinfo{edition}{2nd} edition.
\newblock \bibinfo{publisher}{Routledge}, \doi{10.4324/9780203994627}.

\bibitemdeclare{inproceedings}{rao1995bdi}
\bibitem{rao1995bdi}
\bibinfo{author}{Anand~S. \surnamestart Rao\surnameend} \&
  \bibinfo{author}{Michael~P. \surnamestart Georgeff\surnameend}
  (\bibinfo{year}{1995}): \emph{\bibinfo{title}{BDI agents: From theory to
  practice.}}
\newblock In: {\sl \bibinfo{booktitle}{Proceedings of the First International
  Conference on Multiagent Systems (ICMAS)}}, pp. \bibinfo{pages}{312--319}.

\bibitemdeclare{article}{sebastiani2009automated}
\bibitem{sebastiani2009automated}
\bibinfo{author}{Roberto \surnamestart Sebastiani\surnameend} \&
  \bibinfo{author}{Michele \surnamestart Vescovi\surnameend}
  (\bibinfo{year}{2009}): \emph{\bibinfo{title}{Automated reasoning in modal
  and description logics via SAT encoding: the case study of
  K(m)/ALC-satisfiability}}.
\newblock {\sl \bibinfo{journal}{Journal of Artificial Intelligence Research}}
  \bibinfo{volume}{35}, pp. \bibinfo{pages}{343--389}, \doi{10.1613/jair.2675}.

\bibitemdeclare{article}{silva_1993}
\bibitem{silva_1993}
\bibinfo{author}{Pedro~V. \surnamestart Silva\surnameend}
  (\bibinfo{year}{1993}): \emph{\bibinfo{title}{On the semilattice of
  idempotents of a free inverse monoid}}.
\newblock {\sl \bibinfo{journal}{Proceedings of the Edinburgh Mathematical
  Society}} \bibinfo{volume}{36}(\bibinfo{number}{2}), p.
  \bibinfo{pages}{349–360}, \doi{10.1017/S0013091500018447}.

\bibitemdeclare{article}{silva2000unique}
\bibitem{silva2000unique}
\bibinfo{author}{Pedro~V. \surnamestart Silva\surnameend}
  (\bibinfo{year}{2000}): \emph{\bibinfo{title}{On unique factorization
  semilattices}}.
\newblock {\sl \bibinfo{journal}{Discussiones Mathematicae-General Algebra and
  Applications}} \bibinfo{volume}{20}(\bibinfo{number}{1}), pp.
  \bibinfo{pages}{97--120}, \doi{10.7151/dmgaa.1009}.

\bibitemdeclare{book}{toulmin}
\bibitem{toulmin}
\bibinfo{author}{Stephen~Edelston \surnamestart Toulmin\surnameend}
  (\bibinfo{year}{2003}): \emph{\bibinfo{title}{The Uses of Argument.}}
\newblock \bibinfo{volume}{Updated ed}, \bibinfo{publisher}{Cambridge
  University Press}, \doi{10.1017/CBO9780511840005}.

\bibitemdeclare{inproceedings}{10.1007/11814771_26}
\bibitem{10.1007/11814771_26}
\bibinfo{author}{Dmitry \surnamestart Tsarkov\surnameend} \&
  \bibinfo{author}{Ian \surnamestart Horrocks\surnameend}
  (\bibinfo{year}{2006}): \emph{\bibinfo{title}{FaCT++ Description Logic
  Reasoner: System Description}}.
\newblock In \bibinfo{editor}{Ulrich \surnamestart Furbach\surnameend} \&
  \bibinfo{editor}{Natarajan \surnamestart Shankar\surnameend}, editors: {\sl
  \bibinfo{booktitle}{Automated Reasoning}}, \bibinfo{publisher}{Springer
  Berlin Heidelberg}, \bibinfo{address}{Berlin, Heidelberg}, pp.
  \bibinfo{pages}{292--297}, \doi{10.1007/11814771\_26}.

\bibitemdeclare{book}{van2007dynamic}
\bibitem{van2007dynamic}
\bibinfo{author}{Hans \surnamestart Van~Ditmarsch\surnameend},
  \bibinfo{author}{Wiebe \surnamestart van Der~Hoek\surnameend} \&
  \bibinfo{author}{Barteld \surnamestart Kooi\surnameend}
  (\bibinfo{year}{2007}): \emph{\bibinfo{title}{Dynamic epistemic logic}}.
\newblock \bibinfo{volume}{337}, \bibinfo{publisher}{Springer Science \&
  Business Media}.

\bibitemdeclare{inproceedings}{10.1007/978-3-540-73595-3_38}
\bibitem{10.1007/978-3-540-73595-3_38}
\bibinfo{author}{Christoph \surnamestart Weidenbach\surnameend},
  \bibinfo{author}{Renate~A. \surnamestart Schmidt\surnameend},
  \bibinfo{author}{Thomas \surnamestart Hillenbrand\surnameend},
  \bibinfo{author}{Rostislav \surnamestart Rusev\surnameend} \&
  \bibinfo{author}{Dalibor \surnamestart Topic\surnameend}
  (\bibinfo{year}{2007}): \emph{\bibinfo{title}{System Description: Spass
  Version 3.0}}.
\newblock In \bibinfo{editor}{Frank \surnamestart Pfenning\surnameend}, editor:
  {\sl \bibinfo{booktitle}{Automated Deduction -- CADE-21}},
  \bibinfo{publisher}{Springer Berlin Heidelberg}, \bibinfo{address}{Berlin,
  Heidelberg}, pp. \bibinfo{pages}{514--520},
  \doi{10.1007/978-3-540-73595-3\_38}.

\end{thebibliography}
